\documentclass{llncs}
\usepackage[utf8]{inputenc}

\usepackage{enumitem}

\usepackage[caption=false]{subfig}
\usepackage{blindtext}
\usepackage{graphicx}
\usepackage{booktabs}
\usepackage{amsmath} 
\usepackage{amsfonts} 
\usepackage{amssymb}
\usepackage{wrapfig}
\usepackage[hyphens]{url}

\newcommand{\corr}[1]{#1$^{***}$}
\newcommand{\nzhpage}[0]{protest page}

\begin{document}
%
%
\title{`Dark Germany': Hidden Patterns of Participation in Online Far-Right Protests Against Refugee Housing}

\author{Sebastian Schelter 
\and Jérôme Kunegis\inst{1}}

\institute{
\textsuperscript{1} University of Namur, Belgium}

\maketitle
%
\makeatletter
\makeatother
\begin{abstract}
The political discourse in Western European countries such as Germany has recently seen a resurgence of the topic of refugees, fueled by an influx of refugees from various Middle Eastern and African countries.  
Even though the topic of refugees evidently plays a large role in online and offline politics of the affected countries, the fact that protests against refugees stem from the right-wight political spectrum has lead to corresponding media to be shared in a decentralized fashion, making an analysis of the underlying social and mediatic networks difficult. 
In order to contribute to the analysis of these processes, 
we present a quantitative study of the social media activities of a contemporary nationwide protest movement against local refugee housing in Germany, which organizes itself via dedicated Facebook pages per city. We analyse data from 136 such protest pages in 2015, containing more than 46,000 posts and more than one million interactions by more than 200,000 users. 
In order to learn about the patterns of communication and interaction among users of far-right social media sites and pages, 
we investigate the temporal characteristics of the social media activities of this protest movement, as well as the connectedness of the interactions of its participants. We find several activity metrics such as the number of posts issued, discussion volume about crime and housing costs, negative polarity in comments, and user engagement to peak in late 2015, coinciding with chancellor Angela Merkel's much criticized decision of September 2015 to temporarily admit the entry of Syrian refugees to Germany. Furthermore, our evidence suggests a low degree of direct connectedness of participants in this movement, (i.a., indicated by a lack of geographical collaboration patterns), yet we encounter a strong affiliation of the pages' user base with far-right political parties.
\end{abstract}

\section{Introduction}
In recent years, Europe has experienced a massive influx of refugees from Middle Eastern and African regions, mainly due to civil wars and economic stagnation in these areas. In Germany, this influx peaked in 2015 with 890,000 people seeking asylum~\cite{BMI2016}, an order of magnitude more than the average number in the preceding ten years; in early September of 2015, chancellor Angela Merkel decided to admit the entry of Syrian refugees stuck in South-East European countries. These developments have been accompanied by a steep rise in popularity of German right-wing organizations, especially in the form of the political party \emph{AfD -- Alternative für Deutschland}, (``Alternative for Germany'')~\cite{Arzheimer2016}, which managed to enter the European parliament as well as multiple German state parliaments since its inception in 2012. The AfD and other right-wing organizations successfully leverage social media to communicate with their followers; recent research shows that their gains in popularity are highly correlated with growing interaction rates and user engagement on Facebook~\cite{Schelter2016}. 

The refugees in Germany have been registered and temporarily placed in hastily implemented shelters distributed all over Germany. As a reaction to these refugee shelters (or mere plans to set up refugee housing), a large number of local protest movements have formed against their placement in the corresponding towns. Also, refugee shelters and refugees have become targets of a series of more than a thousand crimes in 2015~\cite{AmadeuAntonio2015}, including arsony of buildings designated for refugees as well as attacks with explosives against inhabitated shelters. Reports indicate that the communication within anti-refugee housing movements happens via dedicated Facebook pages, often titled ``Nein zum Heim'' (``No to the shelter'') or ``\emph{X} wehrt sich'' (``\emph{X} fights back'', where \emph{X} usually stands for the name of a city). Many of these pages promote racist, xenophobic and islamophobic views. Moreover, setting up such Facebook pages to organize protests is explicitly recommended in guidelines published by radical right-wing organizations~\cite{DerDritteWeg2015}, and some of the pages openly advertise such organisations.\footnote{e.g., \url{https://facebook.com/nzh.koepenick/}, \url{https://facebook.com/Kein-Asylheim-in-der-Reinhardt-Kaserne-826300594067680/}, \url{https://facebook.com/Kein-Asylanten-Containerdorf-in-Falkenberg-1497894543825007/}}
To this date, we are not aware of a comprehensive quantitative study of this phenomenon, even though the topic remains at the forefront of German politics as of 2017. To fill this gap, this paper presents a limited quantitative study on the scale of 136 protest pages in 2015, whose contents we have crawled, including more than one million interactions of more than 200,000 users with these pages. Given this data, we focus our efforts on two research questions:
\begin{itemize}
\item[\textbf{RQ~1:}] What are the temporal characteristics of the social media activities of this protest movement?
\item[\textbf{RQ~2:}] What is the degree of connectedness and cooperation in this protest movement?
\end{itemize}
The goal of RQ~1 is to obtain insights into general activity patterns of this protest movement, therefore we analyze summary statistics of the activity on the social media pages. Our main aim is to gain insight into topics and the general keynote of the content posted on these pages, and to find hints on how these activities relate to external events. Our discovered topics closely resemble the topics found in a recent study of German right-wing activism on Twitter~\cite{Puschmann2016}. We observe that several activity metrics such as the number of posts issued, the negative polarity in comments as well as the attraction of new users peak in late 2015. This peak coincides with chancellor Merkel's much criticized decision from early September to temporarily admit the entry of Syrian refugees to Germany~\cite{Walker2015}. 

For RQ~2, we focus on the connectedness of users of the protest pages in order to investigate the nature of cooperation between the participants in this protest movement. We investigate patterns of collaboration and connectedness of the users of protest pages on the level of direct interactions as well as on the level of interactions with the Facebook pages of political parties, which serve as an indicator for indirect connections between the users. Our data points at a low degree of direct connectedness: no regional collaboration patterns between users are apparent and their co-interaction network is highly disconnected. However, our data suggests that the user base of the protest pages is connected on a higher level, as we encounter a strong affiliation with far-right political parties. Interestingly, the affiliation pattern is very similar for both the extremist-right NPD party and the (self-proclaimed) non-extremist conservative AfD party.

The paper is structured as follows.  We first describe the data acquisition methods and the dataset used in Section~\ref{sec:acquisition}.   
In Section~\ref{sec:temporal}, we investigate the temporal characteristics of the movement, and 
in Section~\ref{sec:network}, we analyse its connectedness and cooperation patterns. 
Related work is reviewed in Section~\ref{sec:related}, and
we conclude in Section~\ref{sec:conclusion}. An extended abstract covering parts of this work has been previously published by the authors~\cite{Schelter2017}. 

\section{Data Acquisition}
\label{sec:acquisition}
\begin{wrapfigure}{R}{0.35\textwidth}
   \vspace{-0.9cm}
   \includegraphics[scale=0.35]{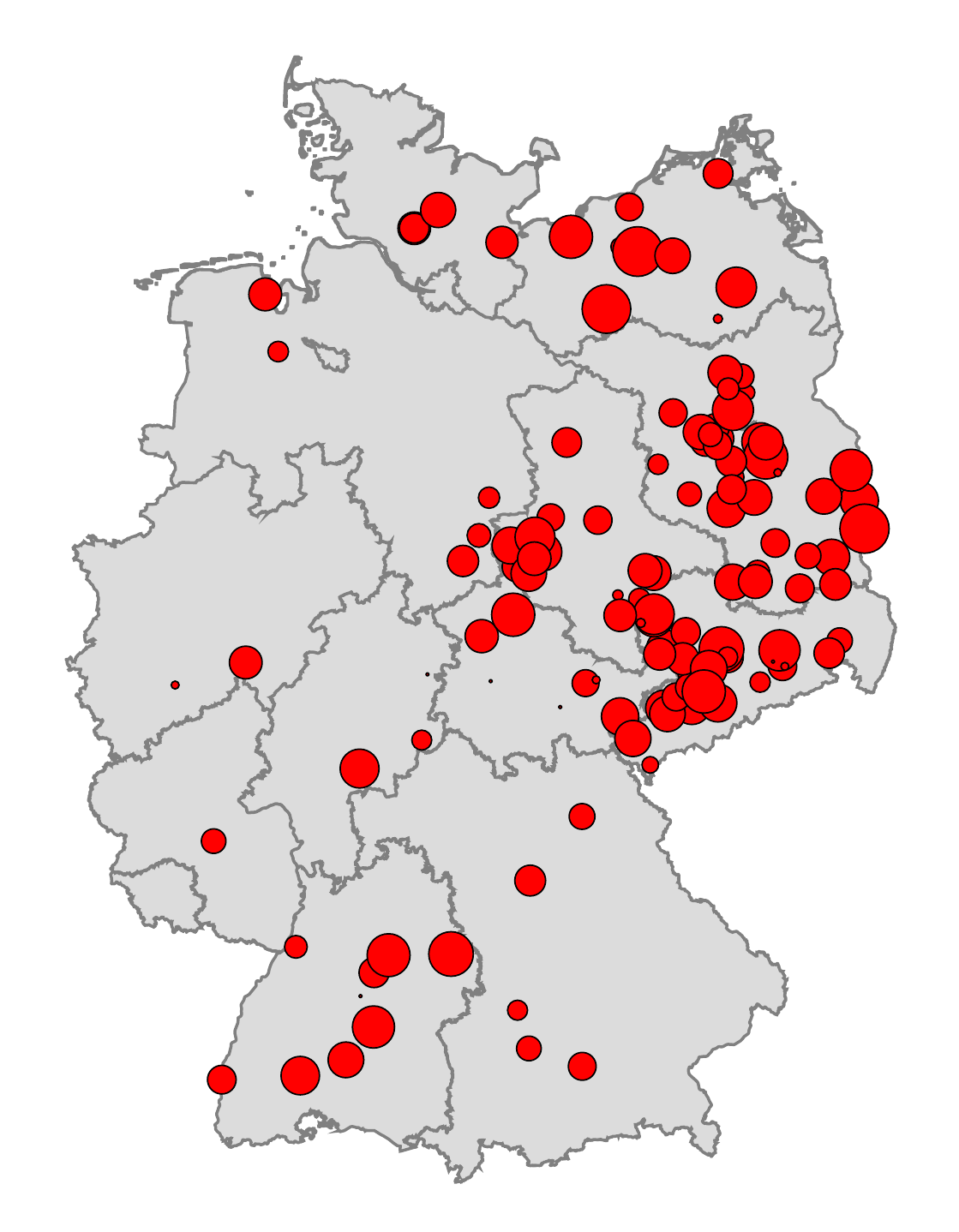}
   \caption{
     Geographical mapping of the pages. The dot size is logarithmically proportional to the number of users of a page. A clustering of pages is apparent in the eastern part of Germany. 
   }
   \label{fig:geo}
   \vspace{-0.8cm}   
\end{wrapfigure}
In order to gather a large number of protest pages for analysis, we consulted online articles listing pages and conducted several exhaustive searches on Facebook using the queries \emph{``Nein zum Heim''} and \emph{``wehrt sich''}, and manually inspected all search results. For the found protest pages, we crawled all publicly available posts with their corresponding likes and comments, restricted to the year 2015. Furthermore, we tagged each \nzhpage\ with the geographical coordinates of the city it refers to. Thereby, we obtained 136 such pages, as depicted in Fig.~\ref{fig:geo}. Our dataset comprises 46,880 posts and more than one million interactions (196,661 comments on posts, 791,072 likes on posts and 339,604 likes on comments) by 209,822 users. We do not have access to shared posts, page-level likes or profile information of users. Note that we only report aggregate statistics of the collected data, and do neither conduct analysis on the level of individual persons, nor re-distribute the data. 

A first look into the data reveals some interesting patterns: As shown in Fig.~\ref{fig:geo} the vast majority (77\%) of protest pages are located in the eastern part of Germany. This matches with the fact that, since the re-unification of Germany in 1990, far-right parties regularly score higher election results in this part of Germany compared to the western regions~\cite{Wikipedia}. We find the distribution of per-page activity metrics (number of posts, users and comments) to be heavily skewed. For instance, the median number of users interacting with a page is 384, while the top 10\% of pages attract more than 5,076 users each, up to a maximum of 30,346. Similarly, the median number of posts per page is 126, while the top 10\% of pages issued more than 780 posts with a maximum of 5,643. The same also holds for comments, where the median is 213, the top 10\% pages see more than 3,476 comments up to a maximum of 24,866.

\section{RQ 1: Temporal Characteristics}
\label{sec:temporal}
As are any political and other movements, the anti-refugee protest
movement is subject to temporal fluctuations -- both due to internal and
external changes.  In order to understand such fluctuations in the community, 
we investigate a wide range of activities, such as posts and news articles issued by pages, the topics in these news articles, as well as the sentiment in user comments and general user engagement. We intend to gain insights into the general keynote of the content posted on the protest pages, and to find hints on how these activities relate to external events.  Specifically, we investigate four aspects:  (1)~the general evolution over time of the volume of activity, (2)~the change in the topics discussed and shared, (3)~changes in the polarity of interactions, i.e., by positive and negative discourse, and (4)~the ability of individual pages to attract a new audience over time. 

\begin{figure}[b!]
  \centering
  \includegraphics[scale=0.4]{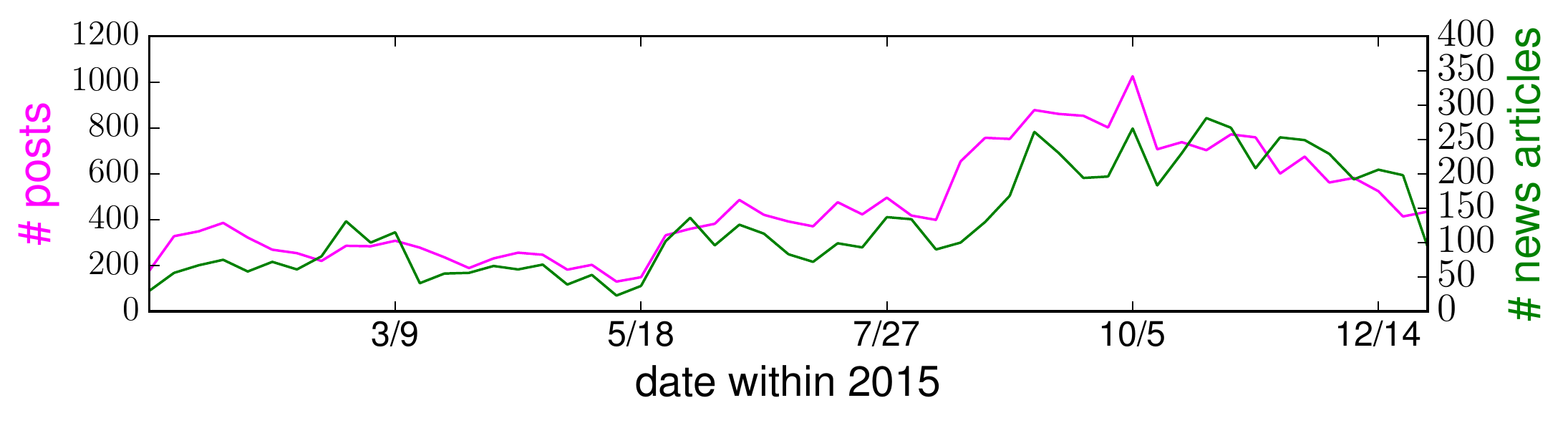}
  \caption{
    Weekly aggregated time course of the number of posts and posted news articles on all protest pages in 2015.  The peak in September/October follows the announcement of admission of Syrian refugees into Germany. 
  }
  \label{fig:posts-with-entites}
\end{figure}

\subsection{Time Course of Page Posts}
We analyze the time course of the number of published posts to gain insight into a general activity pattern of the pages. Fig.~\ref{fig:posts-with-entites} shows the weekly number of posts as issued by the pages and the corresponding number of posted news articles over time. We observe a peak in the end of the third quarter of 2015, which coincides with the aforementioned admission of Syrian refugees into Germany in September~\cite{Walker2015}. The same phenomenon has also been recognized in previous studies of far-right engagement on social media~\cite{Schelter2016}.


\subsection{Topics over Time in Posted News Articles}
Next, we analyze the contents of the posted news articles to gain insights into the conversation on the pages, as well as their time course. Therefore, we crawl the contents of news articles linked to in the posts on the protest pages. We clean the resulting textual data, and represent every news article as a bag-of-words of its nouns and named entity terms, extracted via a part-of-speech tagger~\cite{Toutanova2003}. The resulting dataset comprises 6,760 articles, 7,548,572 word tokens, and 17,402 distinct terms. In order to investigate the topics eminent in these articles, we conduct topic analysis via a variant of Latent Dirichlet Allocation (LDA) called \emph{Topics over Time}~\cite{Wang2006} that not only captures the low-dimensional structure of post contents, but also how this structure changes over time. We select the number of topics via manual inspection of the resulting topic clusters, to maximize the interpretability of the model. Finally, we fit a model with ten clusters, employing the default hyperparameter settings of the implementation, which the authors report to be robust among several datasets.

\begin{table*}
  \centering
   \caption{
     Selection of topics from a time-sensitive topic model of news articles in posts on the protest pages. For each selected topic, we provide a manually chosen label, illustrate its time course via a histogram of the estimated document-topic matrix (binned weekly by posting dates of the articles), and list the five most likely terms, as well as headlines of a set of strongly associated news articles.  Terms and headlines are translated from German. 
   }    
  \begin{tabular}{p{6.1cm} p{6.1cm}}
    \textbf{Housing and Cost} & \textbf{Sexual Crime} \\
    \midrule
   \includegraphics[scale=0.4]{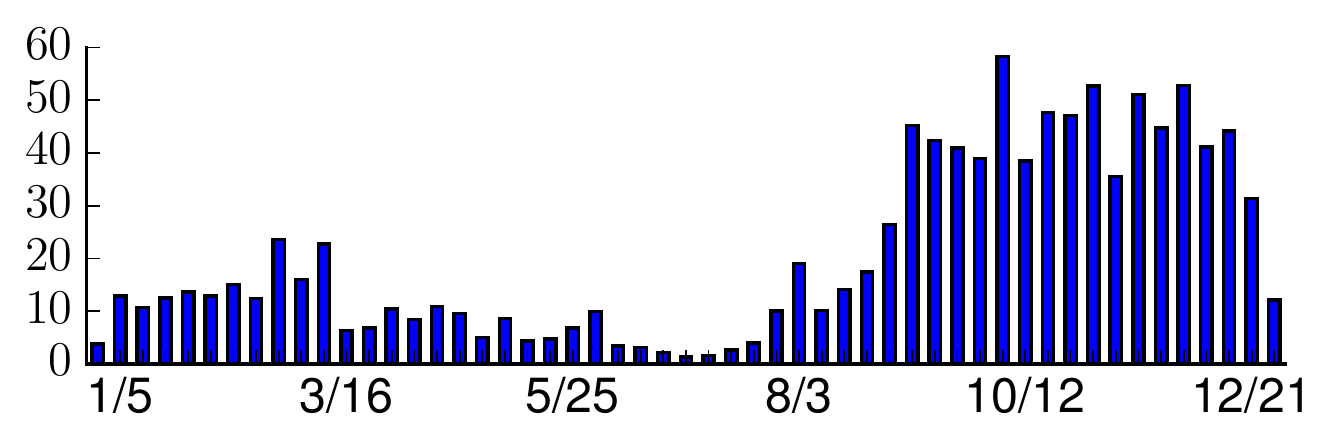} & 
   \includegraphics[scale=0.4]{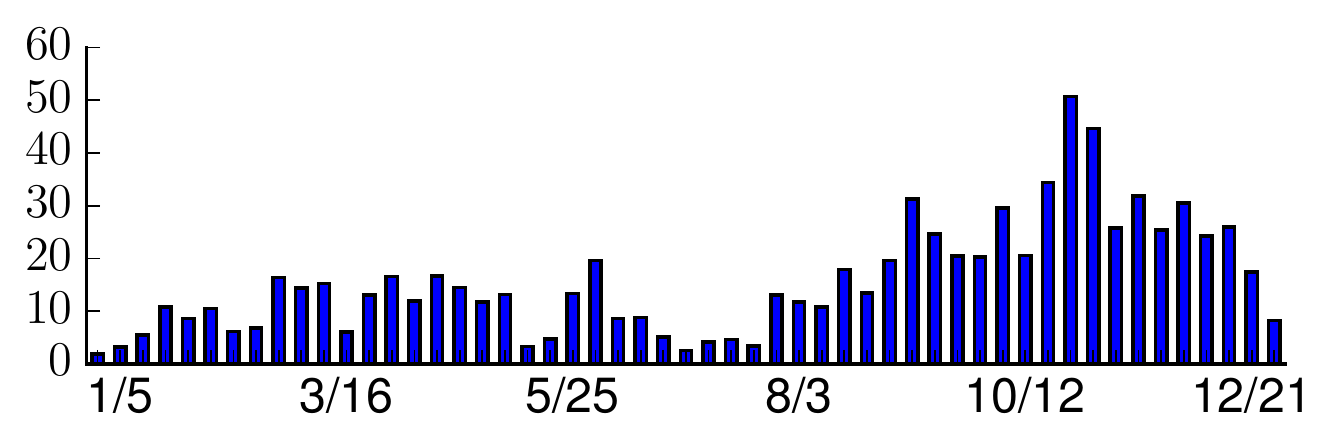}\\[-0.1cm]

   \midrule

    \begin{minipage}{.33\textwidth}
      \begin{tabular}{p{4.5cm} p{.6cm}}
        {\em refugees} & 0.01251 \\    
        {\em asylum seeker} & 0.00941 \\
        {\em refugee} & 0.00736 \\
        {\em city} & 0.00728 \\
        {\em housing} & 0.00648 \\
      \end{tabular}  
    \end{minipage}
    & 
    \begin{minipage}{.33\textwidth}
      \begin{tabular}{p{4.5cm} p{.6cm}}
        {\em man} & 0.01146 \\
        {\em perpetrator} & 0.00881 \\   
        {\em years} & 0.00698 \\
        {\em years old} & 0.00691 \\  
        {\em woman} & 0.00651 \\
     \end{tabular}
   \end{minipage}
  \\
  \midrule

  \begin{minipage}{.33\textwidth}
    \begin{tabular}{p{4.3cm} p{.9cm}}
      {\em ``Dresden invests 47.7 million euros for asylum seekers''} & $\;\;\;\;\;\;\;$0.94\\ 
      {\em ``County builds new shelters for refugees''} & $\;\;\;\;\;\;\;$0.93 \\ 
      {\em ``Welcome to New-Aleppo, the refugee city''} & $\;\;\;\;\;\;\;$0.91 \\ 
    \end{tabular}
  \end{minipage}

  & 

  \begin{minipage}{.22\textwidth}
    \begin{tabular}{p{4.3cm} p{.9cm}}
      {\em ``29-year old woman sexually harassed by unknown man''} & $\;\;\;\;\;\;\;$0.97\\ 
      {\em ``Call for witnesses after rape in Dresden-Plauen''} & $\;\;\;\;\;\;\;$0.95  \\ 
      {\em ``Another sexual attack on a young woman in Dresden''} & $\;\;\;\;\;\;\;$0.94 \\ 
    \end{tabular}
  \end{minipage}\\[1cm]

  & \\

  \textbf{Europe} & \textbf{Conflict with the Left} \\
  \midrule
   \includegraphics[scale=0.4]{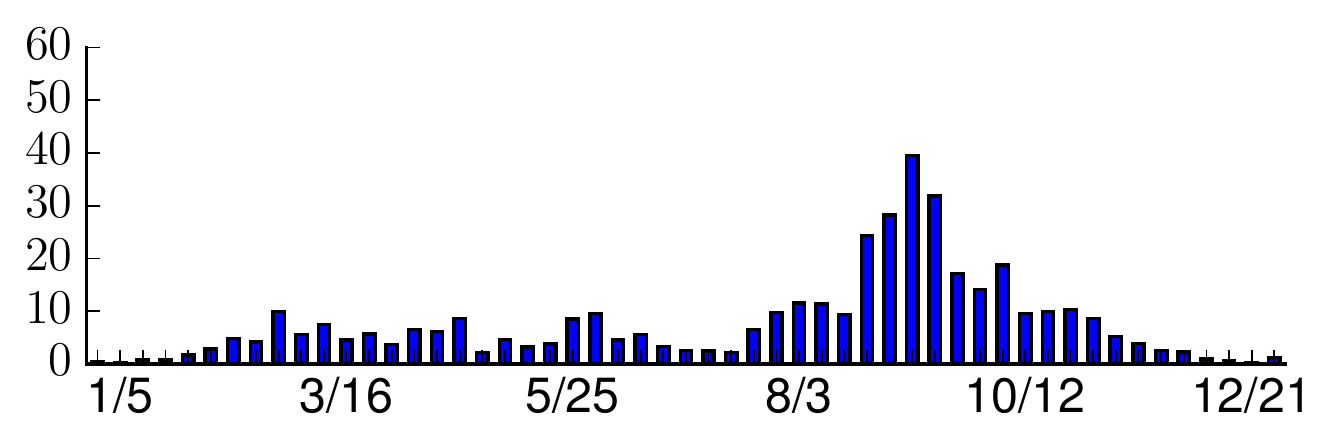} & 
   \includegraphics[scale=0.4]{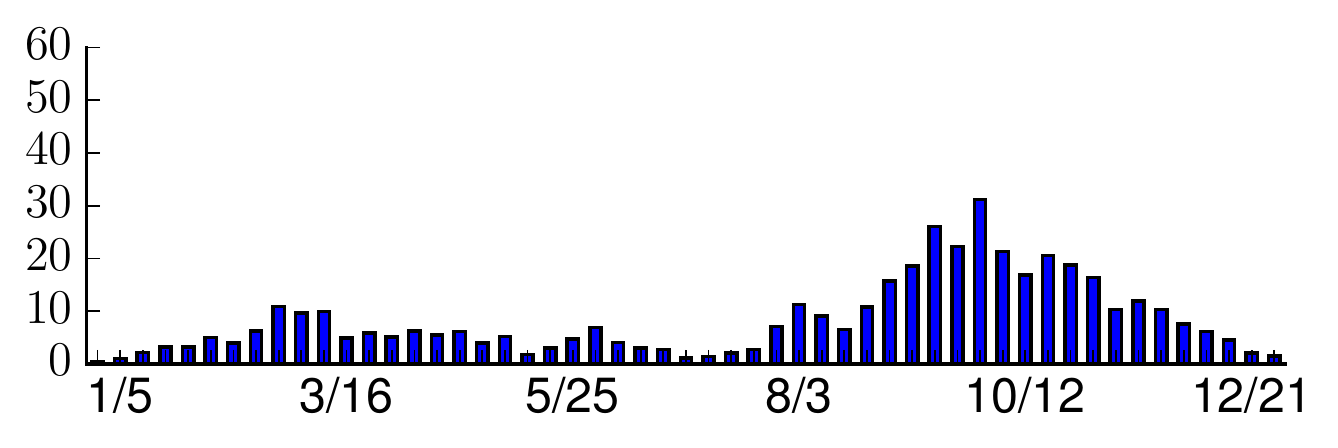}\\[-0.1cm]
  \midrule 

    \begin{minipage}{.33\textwidth}
      \begin{tabular}{p{4.5cm} p{.6cm}}
      {\em refugees} & 0.00917 \\
      {\em germany} & 0.00669 \\
      {\em hungary} & 0.00604 \\
      {\em eu} & 0.00572 \\
      {\em humans} & 0.00556 \\
     \end{tabular}
   \end{minipage}  

 & 

    \begin{minipage}{.33\textwidth}
      \begin{tabular}{p{4.5cm} p{.6cm}}
       {\em article} & 0.00857 \\
       {\em donation} & 0.00718 \\
       {\em support} & 0.00535 \\
       {\em humans} & 0.00523 \\
       {\em place} & 0.00501 \\                                                          
     \end{tabular}
   \end{minipage}\\

   \midrule 

  \begin{minipage}{.22\textwidth}
    \begin{tabular}{p{4.3cm} p{.9cm}}
       {\em ``The ruins of asylum policy''} & $\;\;\;\;\;\;\;$0.95 \\ 
       {\em ``Hungary closes route over the Balkans''} & $\;\;\;\;\;\;\;$0.77 \\ 
       {\em ``Germany is Europe's refugee shelter''} & $\;\;\;\;\;\;\;$0.73 \\ 
     \end{tabular}
   \end{minipage}  

 &

  \begin{minipage}{.22\textwidth}
    \begin{tabular}{p{4.3cm} p{.9cm}}
       {\em ``Federal police chief speculates about leftist perpetrators after arsony in Tröglitz''} & $\;\;\;\;\;\;\;$0.90\\ 
       {\em ``Antifacists scare off Saxony's minister of the interior''} & $\;\;\;\;\;\;\;$0.84 \\ 
     \end{tabular}
  \end{minipage}\\
  \midrule
  &\\
  \end{tabular}
   \label{tab:topics}  
\end{table*}  

For each topic, we report the five terms (translated from German) with highest likelihood of occurring in the topic, as well as the (translated) headlines of a set of news articles strongly associated with the topic. Analogously to~\cite{Wang2006}, we illustrate the time course of the topics via a histogram of the document-topic assigment matrix learned by Latent Dirichlet Allocation, where the binning is based on the weeks of the articles' post date. Finally, we manually choose a label for each topic by inspecting the corresponding most likely terms and documents. We list data for four topics in Table~\ref{tab:topics}. 

The first topic we identify comprises discussions about housing capacities and costs, indicated by terms such as \textit{city} and \textit{housing}, and strongly associated news articles that talk about the rising costs for cities that set up housing for refugees. A second topic is concerned with sexual crimes, indicated by its most likely terms such as \textit{man}, \textit{perpetrator}, and \textit{woman}, and by the headlines of strongly associated news articles concerning sexual violence offenses. These two topics seem to be general motives of the right-wing agenda, as both of these grow relatively constant with the volume of posted articles over time. A topic that peaks in September 2015 comprises discussions about the fact that the majority of European countries did not follow Germany's example in admitting the entry of large numbers of refugees, despite requests of the German government. This topic is indicated by terms such as \textit{germany}, \textit{hungary} and \textit{eu}, and news articles that discuss that countries such as Hungary opposed German political demands for taking in refugees by closing their intra-European borders to Croatia in October 2015. Lastly, we encounter a topic that reflects the conflicts with leftist activists, indicated by terms such as \textit{donation} or \textit{support} and news articles that talk about militant action attributed to antifascist groups. The latter topics appear to be niche topics, which peak at special events, such as Hungary closing its borders or standoffs between right-wing and leftist activists in August 2015. 
~\\
~\\
~\\

\begin{figure}[b]
  \centering
  \includegraphics[scale=0.6]{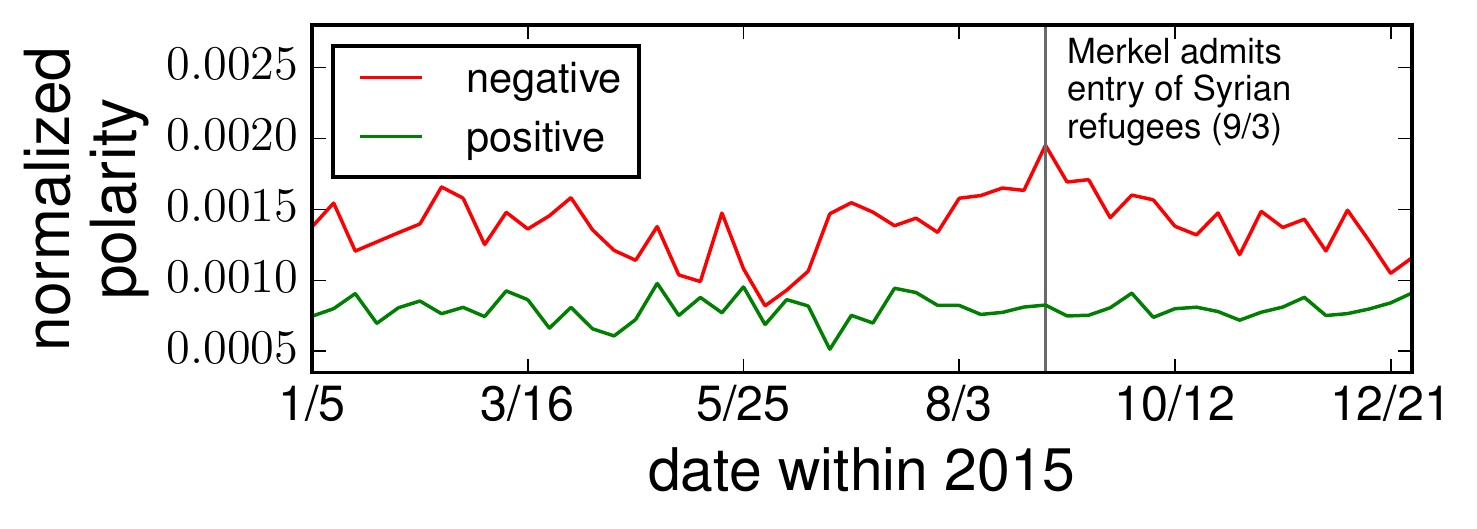}
  \caption{
    Time course of normalized negative and positive polarity in user comments per week in 2015, computed from a sentiment dictionary.  The vertical line marks chancellor Angela Merkel's decision to admit Syrian refugees. 
  }
  \label{fig:polarity}
\end{figure}

\subsection{Polarity in User Comments}
Next, we place our focus on the users interacting with the pages and investigate the time course of overall sentiment in the user comments. For that, we employ a dictionary denoting the negative sentiment $\phi^{-}(t)$ and positive sentiment  $\phi^{+}(t)$ of a German language term $t$ when used in certain parts of speech~\cite{Waltinger2010a}. We apply part-of-speech tagging to all comments in a given week $w$, which gives us all contained terms $T_w$. Next, we sum the contained polarity using the weights of the aforementioned dictionary and normalize the results by the number of terms in the comments per week to compute the normalized negative polarity $p_{w^-}$ 
and normalized positive polarity $p_{w^+}$ per week 
\begin{align*}
  p_{w^{\pm}} = \frac{1}{|T_w|} \sum_{t \in T_w} \phi^{\pm}(t). 
\end{align*}
Fig.~\ref{fig:polarity} shows the resulting time course of these polarities for all weeks in 2015. We observe that negative speech dominates the comments throughout the whole year. Furthermore, we encounter a peak in negative polarity, which again coincides with chancellor Merkel's decision from early September 2015 to admit the entry of Syrian refugees to Germany~\cite{Walker2015}. This finding confirms the general perception that Merkel's decision provoked widespread anger in the far-right political spectrum.

\subsection{User Attraction}
\begin{figure}[b!]
  \centering
  \includegraphics[scale=0.6]{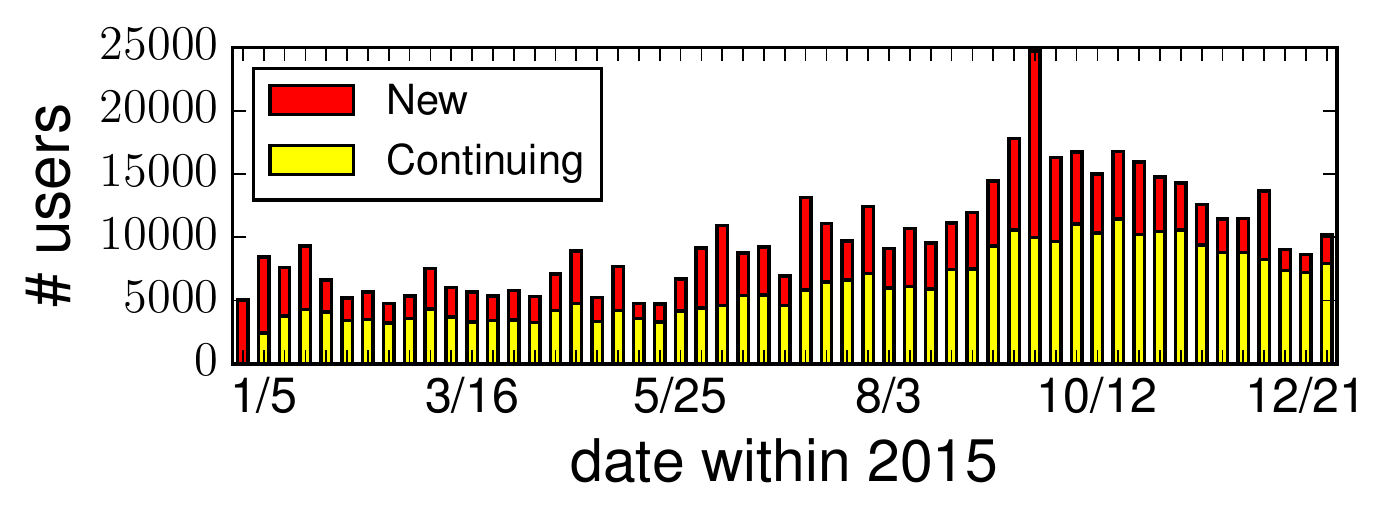}
  \caption{Stacked bar plot of the time course of weekly active users on the protest pages in 2015. We distinguish between new users (active for the first time) and continuing users.}
  \label{fig:new-and-continuing-users}
\end{figure}

Finally, we analyze the pages' ability to attract users over time. For that, we compute the set of active users $U_w$ for every week $w$,~(users who comment on a post or like a post on at least one of the protest pages during that week). For every week $w$, we split these active users into two groups: \emph{new} users 
\begin{align*}
  U_{w_{\mathrm{new}}} = \{ u \mid u \in U_w \wedge u \notin U_v \forall v \in 0, \dots, w-1 \},
\end{align*}
which we encounter for the first time and \emph{continuing} users $U_{w_{\mathrm{cont}}} = U_w \setminus U_{w_{\mathrm{new}}}$, whom we have already seen previously. The corresponding sizes of these users sets for all weeks in 2015 are shown in Fig.~\ref{fig:new-and-continuing-users}. We see a slight increase in both new and continuing users in the late second half of 2015. However, this increase starts to diminish again towards the end of the year. We find that the time series of continuing users is very strongly correlated with the time series of posts~(\corr{0.91})\footnote{$\;$ \corr{} indicates significance at the $p <0.001$ level} and comments~(\corr{0.87}) in the corresponding week. We observe a similarly directed but much weaker correlation between the time series of posts and new users (\corr{0.55}) and posts and comments (\corr{0.56}). Furthermore, we note that the number of mean weekly active users~(9,935) is very small compared to the overall number of users. We encounter a strong correlation~(\corr{0.87}) of the number of continuing users with time index~$i$, but cannot determine a similar significant correlation for new users. These findings suggest that the protest pages maintain a low constant growth of users, but fail at accelerating this growth.

\section{RQ 2: Connectedness and Cooperation}
\label{sec:network}
The German anti-refugee movement is inherently decentralized -- there is no dominating ``no to refugees'' or similar page that attracts the majority of likes, as is the case for a large number of non-controversial topics on Facebook.  As a result, the likes are spread among a much larger number of pages, which ostensibly follow a geographical pattern, i.e., most such pages are, at least by name, specific to a single city or small region. 
In order to investigate to what extent the underlying social network itself exhibits collaboration patterns, we therefore shift our focus from the contents and time course of activities to the relationships between the participating users. In particular, we investigate patterns of collaboration and connectedness of the protest page users on two levels: We first investigate direct co-interactions between users on the social networking platform itself, and in a second step, we investigate their affiliation with the Facebook pages of political parties. This affiliation serves as an indicator for indirect connections between the users, in particular since individual friendship links are not made crawlable by Facebook. These investigations can act as input to future research on the longevity of the currently observed rise of far-right movements.
In the following, we investigate (1)~whether the geographical distance between two pages is reflected in the shared user base between the pages, (2)~the presence of a giant connected component in the co-like network, and (3)~the degree of affiliation for Facebook pages of right-wing political pages.

\subsection{Low Correlation between Geographical Distance and Amount of Shared Users}
In order to investigate geographical aspects of the data, we compute the geographical distance between the corresponding cities for each pair of pages, and compare this to the Jaccard similarity between their sets of users. We would expect to see a strong negative correlation if geographical closeness implicated co-operating user bases. However, the maximum Jaccard similarity is only 0.1428, and 4,727 page pairs exhibit non-zero similarity, leaving 3,916 page pairs with zero shared users. Even for pairs of pages with non-zero similarity, the correlation is low (\corr{$-$0.19}), which suggests against a geographical collaboration pattern. 

\subsection{Absence of a Giant Connected Component in the User Co-like Network}
Next, we construct the \emph{user co-like network} as follows: users form the vertices of this network, and for every post, we introduce edges between all users that liked the post. The resulting network has 95,639,173 edges (co-likes among users). We study the connectivity of this network by computing the size of its largest connected component. This size amounts to 89,094 users, which account for only 57.5\% of the overall user base. This is very atypical for real-world social networks, which typically exhibit a giant connected component containing nearly all users,\footnote{See e.g.\ \url{http://konect.uni-koblenz.de/statistics/coco}} and gives a hint that the social media activities of the users on the pages might be highly separated.  This is surprising in light of the fact that the users, as found in the previous experiment, were found to not follow a clustering into regions, hinting that another, non-geographic clustering of users is present in the data.  The present dataset however does not allow us to identify the nature of this clustering. 

\subsection{Strong Affiliation with Far-right Organizations}
As we could not find evidence for collaboration patterns in the direct interactions between users, we investigate whether the users of the pages are connected on a higher level. Therefore, we analyse the affiliations of the users on these protest pages to political parties in Germany, with the aim to see whether these users are connected in that way. For that, we employ additional data about likes of posts on the parties' Facebook pages from our previous work~\cite{Schelter2016}.
Next, we compute the affiliation $\text{aff}_{p,o}$ between a page $p$ and a political party $o$ as the ratio of users~$U$ interacting with the page that also liked posts on the party's page: 
\begin{align*}
\text{aff}_{p,o} = |U_{\text{interact-with}(p)} \cap U_{\text{like-post}(o)}| \; / \; |U_{\text{interact-with}(p)}|
\end{align*}
In the resulting distributions, we observe that the median affiliation with the right-wing parties \emph{AfD} (0.45) and \emph{NPD} (0.41) is about one order of magnitude higher than the affiliation with parties from the remaining political spectrum, such as the christian-conservative \emph{CDU}~(0.04), the social-democratic \emph{SPD}~(0.04), the green party \emph{Die Grünen}~(0.03), and the socialist-left party \emph{Die Linke}~(0.02). 
While it is expected to see a strong affiliation with the \emph{NPD}, which is commonly considered to be the voice of the extreme right and has repeatedly been the target of party-ban trials by the German state, it is suprising to see an even stronger affiliation with the \emph{AfD}, as the latter party claims to locate itself in the conservative spectrum rather than the extremist-right spectrum.

\section{Related Work}
\label{sec:related}
The social media usage of political movements is of interest to many studies, e.g., with a focus on the \emph{Black Lives Matter} movement in the United States~\cite{Choudhury2016,Olteanu2016}. However, the German far-right has seen little attention so far, with current research mostly focusing on exploratory analysis of the social media activities of the AfD party on Facebook~\cite{Schelter2016} and the topics discussed by the local anti-immigrant movement \emph{Pegida}~\cite{Puschmann2016} on Twitter, as well as their corresponding news sources~\cite{Puschmann2016info}. The concentration on social concerns with crime and housing cost, and the focus on European policies in topic clusters discovered by Puschmann et al.\ on Twitter closely resemble the topics we discovered in Section~\ref{sec:temporal}.

The rise of populist radical right parties in Europe has been extensively researched. While other countries, such as France, Denmark, Belgium, and Austria have experienced high levels of voting for populist radical right parties, for a long time Germany seemed to have been an exception with regard to the radical right in Western Europe~\cite{Arzheimer2016}. With the rise of the AfD party, Germany now also becomes part of the ``pathological normalcy''~\cite{Mudde2010}. The term denotes that populist right-wing ideology is a radicalization of mainstream values, such as ethnic nationalism, anti-immigrant sentiment and authoritarian values. These attitudes have been present within a segment of the population even before the rise of the AfD, but have not been represented within mainstream party politics before. Scholarship on the conditions of radical right success have associated it with a convergence of the mainstream parties on the left and right, leaving a representational gap for the radical right to move in~\cite{Arzheimer2006,Kitschelt1997,Van2005}, while some are sceptical of this ``convergence of the middle'' thesis~\cite{Norris2005}.

\section{Conclusion}
\label{sec:conclusion}
We studied the Facebook activities of a contemporary nationwide protest movement against refugee housing in Germany. We analysed data from 136 public Facebook pages, containing more than one~million interactions by more than 200,000 users. We encountered peaks in several activity metrics that coincide with chancellor Merkel's decision to temporarily admit the entry of Syrian refugees to Germany, which suggests that this political move caused anger and outrage in far-right circles. However, despite the presumed mobilization effects stemming from Merkel's policies in 2015, our evidence suggests a low degree of user growth, connectedness and cooperation in this protest movement. From all German political parties, the {\em AfD} exhibited the strongest affiliation among the user base of the studied protest pages, which contradicts previous classifications of the {\em AfD} as not belonging to the far-right political spectrum~\cite{Arzheimer2016}. 
Furthermore, while we can confirm that the Facebook pages of the anti-refugee movement in Germany are split into many small pages by geography, with no apparent regional collaboration patterns. 
A noteworthy limitation of our work is the lack of data on comparable movements on Facebook, which would allow for stronger conclusions about the specifity of the observed trends for the German radical right. In future work, we plan to conduct a deeper analysis of textual contents of user comments, in order to be able to measure controversy on the level of user interactions.  

\section*{Acknowledgments}
This research was partly funded by the European Regional Development Fund (ERDF/FEDER -- IDEES).   

\bibliographystyle{splncs03}
\end{document}